\documentclass[aps,prd,preprintnumbers,showpacs]{revtex4}
\usepackage{graphicx}

\begin{document}

%
%

\eprint{Nisho-2-2025}
\title{A Way of Axion Detection with Mass $10^{-4} \text{-}10^{-3}$eV Using Cylindrical Sample with Low Electric Conductivity}
\author{Aiichi Iwazaki}
\affiliation{International Economics and Politics, Nishogakusha University,\\ 
6-16 3-bantyo Chiyoda Tokyo 102-8336, Japan }   
\date{Oct. 19, 2025}
\begin{abstract}
A dark matter axion with mass $m_a$ induces an oscillating electric field in a cylindrical sample placed under a magnetic field $B_0$ parallel to the cylinder axis. When the cylinder is made of a highly electrically conductive material, the induced oscillating current flows 
only at the surface.
In contrast, if the cylinder is composed of a material with small conductivity, e.g. $\sigma = 10^{-3}\text{eV}$, the electric
current flows inside the bulk of the cylinder.
Within the QCD axion model, the current $I$ is estimated as
$I(\sigma=10^{-3}\text{eV})\simeq 2.8\times 10^{-14}\text{A}g_{\gamma}\big(R/6\text{cm}\big)^2
\big(\sigma/10^{-3}\text{eV}\big)\big(B_0/15\text{T}\big)\big(10/\epsilon\big)\big(\rho_a/0.3\rm GeVcm^{-3}\big)^{1/2}$ for $m_a=10^{-4}$eV, with radius $R$, permittivity $\epsilon = 10$ of the cylinder and axion energy density $\rho_a$, where $g_{\gamma}$ is
model dependent parameter;
$g_{\gamma}(\text{KSVZ}) = -0.96$ and $g_{\gamma}(\text{DFSZ}) = 0.37$.
Because the current is proportional to $R^2$, using large sample with $R=80$cm,
we have large signal-noise ratio ( $>1$ ) even in temperature $T=4$K,
$I(\sigma=10^{-3}\text{eV})/I_n({\sigma=10^{-3}\text{eV})}\times \sqrt{\delta \omega \delta t_{ob}/2\pi} 
\simeq  
1.1g_{\gamma}(4\text{K}/T)^{1/2}(L/100\text{cm})^{1/2}(R/80\text{cm})
(B_0/7\mbox{T})(\rho_a/0.3\rm GeVcm^{-3})^{1/2}
(\delta t_{ob}/10^3\,\text{s})^{1/2}$ for $m_a=10^{-4}\text{eV}$
with $\epsilon=10$ and $\sigma=\epsilon m_a$,
where thermal noise is $I_n=\sqrt{2T\delta \omega/\pi R_c}$ with $\delta \omega=10^{-6}m_a$
and resistance $R_c=L/(\sigma \pi R^2)$ of the cylinder with length $L$.
Although a superconducting solenoid sufficiently large to accommodate such a sample is required,
the detection of dark matter axions in our proposal may be feasible in the mass range $m_a =10^{-4}\text{-}10^{-3}\text{eV}$.

\end{abstract}
\hspace*{0.3cm}

\hspace*{1cm}

\maketitle

\section{introduction}
A central issue in particle physics is to identify phenomena beyond the Standard Model. The axion, proposed as the Nambu–Goldstone boson of Peccei–Quinn symmetry \cite{axion,axion1,axion2}, provides a natural solution to the strong CP problem and is also a well-motivated dark matter candidate. It is called as QCD axion.
The viable mass window for the QCD axion is tightly constrained to $m_a = 10^{-6}\mbox{-}10^{-3}\text{eV}$ \cite{Wil,Wil1,Wil2}.

Numerous experiments are underway to search for dark matter axions \cite{review}, most exploiting axion–photon conversion in a strong magnetic field. The induced electromagnetic radiation is expected to be
detected with resonant cavities\cite{admx,haystac}, superconducting qubits\cite{moroi}, Quantum Hall effect \cite{iwazaki,aiwazaki}, e.t.c.

\vspace{0.1cm}
In this letter we propose a new method for axion detection using a cylindrical sample with small electrical conductivity, 
$\sigma = 10^{-3}\text{-}10^{-2}\text{eV}$ at low temperature $\sim 1$K. Our target is 
the mass range $m_a=10^{-4}\mbox{-} 10^{-3}$eV, which is difficult to
detect with resonant cavity.

Dark matter axion generates an oscillating electric field in the presence of a strong magnetic field, which in turn induces an oscillating current in the cylinder. By applying the magnetic field parallel to the cylinder axis, the induced current flows parallel to the external field.

In general, such currents are confined within the skin depth $\delta$. For highly conductive materials, e.g. $\sigma = 10^4\text{eV}$, the skin depth is extremely small ( $\delta \sim 10^{-5}\text{cm}$ for $m_a=10^{-4}$eV ), restricting the current to a thin surface layer
due to the induction effect. 
Additionally, electric field induced in the conducting material is suppressed by the factor $\sqrt{m_a/\sigma}\sim 10^{-4}$, compared with
the one induced in vacuum.  
As a result, the electric current is not large enough to be detectable.

On the other hand, the two suppression factors do not arise in materials with small conductivity such as semiconductor. 
One is that we have no suppression factor, $\sqrt{m_a/\sigma}$ 
for the cylinder sample with small conductivity $\sigma=10^{-3}$eV. The other one is that
the induced current mainly flows the bulk. 
The presence of the bulk current is a distinctive feature of electromagnetic fields coupled with axion. 
The resulting electric current $I$ 
and power $P$ become large with large sample. It would be detectable at low temperatures ( $T \sim 1\text{K}$ ).

We show that when the axion mass $m_a=10^{-4}$eV,

\begin{eqnarray}
I(\sigma=10^{-3}\text{eV})&\simeq& 2.9\times 10^{-14}\text{A}g_{\gamma}\Big(\frac{R}{6\text{cm} }\Big)^2
\Big(\frac{B_0}{15\text{T}}\Big)\Big(\frac{2\big(\frac{\epsilon}{10}\big)\big(\frac{\sigma}{10^{-3}\text{eV}}\big)\big(\frac{m_a}{10^{-4}\text{eV}}\big)^2}{\big(\frac{\epsilon}{10}\big)^2\big(\frac{m_a}{10^{-4}\text{eV}}\big)^2+\big(\frac{\sigma}{10^{-3}\text{eV}}\big)^2}\Big)
\Big(\frac{\rho_a}{0.3\rm GeVcm^{-3}}\Big)^{1/2} \\
P(\sigma=10^{-3}\text{eV})&\simeq& 5.3\times 10^{-27}\mbox{W}g_{\gamma}^2
\Big(\frac{L}{100\mbox{cm}}\Big)\Big(\frac{R}{6\mbox{cm}}\Big)^2
\Big(\frac{B_0}{15\mbox{T}}\Big)^2\Big(\frac{2\big(\frac{\sigma}{10^{-3}\text{eV}}\big)\big(\frac{m_a}{10^{-4}\text{eV}}\big)^2}{\big(\frac{\epsilon}{10}\big)^2\big(\frac{m_a}{10^{-4}\text{eV}}\big)^2+\big(\frac{\sigma}{10^{-3}\text{eV}}\big)^2}\Big)
\Big(\frac{\rho_a}{0.3\rm GeVcm^{-3}}\Big)
\end{eqnarray}
with the choice $\sigma=10m_a=10^{-3}$eV,
where $R ( L )$ denotes the radius ( length ) of the cylinder with permittivity $\epsilon=10$.  
The power is maximized by choosing electrical conductivity such as $\sigma =\epsilon m_a$, 
for instance, $\sigma=10^{-3}\text{eV}(m_a/10^{-4}\text{eV}) $ when $\epsilon=10$.
We note that $I$ and $P(\sigma=10^{-3}\text{eV})$ are proportional to the surface area $R^2$.

The oscillating current flows the entire cylindrical sample, 
in contrast to ordinal metal with large conductivity, in which electric currents with high frequencies are confined to the surface
due to electromagnetic induction. 

Indeed, when the conductivity is large, for instance, $\sigma=10^4$eV, the power is given by

\begin{equation}
P(\sigma=10^4\text{eV}) \simeq 5.4\times 10^{-31}\mbox{W}g_{\gamma}^2\Big(\frac{L}{100\mbox{cm}}\Big)\Big(\frac{R}{6\mbox{cm}}\Big)
\sqrt{\frac{m_a}{10^{-4}\mbox{eV}}}\sqrt{\frac{10^4\mbox{eV}}{\sigma}}
\Big(\frac{B_0}{15\mbox{T}}\Big)^2\Big(\frac{\rho_a}{0.3\rm GeVcm^{-3}}\Big)
\end{equation}

Obviously, the power $P(\sigma=10^4\text{eV})$ proportional to $R$ 
is fourth order of magnitude smaller than $P(\sigma=10^{-3}\text{eV})$.

\vspace{0.2cm}
In the observation,
the electric current $I(\sigma=10^{-3}\text{eV})$ should be compared with thermal noise $ I_n=\sqrt{2T\delta \omega/\pi R_c}$ where 
the resistance $R_c=L/(\sigma \pi R^2)$ of the cylinder and $\delta \omega=10^{-6}m_a$,

\begin{eqnarray}
\label{4}
\frac{I\big(\sigma=(10^{-3}\text{eV}/\sqrt{3}\big))}{I_n(\sigma=\big(10^{-3}\text{eV}/\sqrt{3})\big)}&\simeq& 5.0\times 10^{-4}g_{\gamma}\Big(\frac{20\text{mK}}{T}\Big)^{1/2}
\Big(\frac{L}{100\mbox{cm}}\Big)^{1/2}
\Big(\frac{R}{6\text{cm}}\Big)\nonumber \\
&\times &\Big(\frac{B_0}{15\text{T}}\Big)\Big(\frac{2\big(\frac{\epsilon}{10}\big)\big(\frac{\sqrt{3}\sigma}{10^{-3}\text{eV}}\big)^{1/2}\big(\frac{m_a}{10^{-4}\text{eV}}\big)^{3/2}}{\big(\frac{\epsilon}{10}\big)^2\big(\frac{m_a}{10^{-4}\text{eV}}\big)^2+\big(\frac{\sqrt{3}\sigma}{10^{-3}\text{eV}}\big)^2}\Big)
\Big(\frac{\rho_a}{0.3\rm GeVcm^{-3}}\Big)^{1/2},
\end{eqnarray}
with $m_a=10^{-4}$eV,
where we have taken $\sigma=\epsilon m_a/\sqrt{3}=(10^{-3}\text{eV}/\sqrt{3})$, which maximizes the ratio $I/I_n$.  

The noise is further reduced by considering a frequency width $\delta \omega=10^{-6}m_a$ 
and an observation time $\delta t_{ob}=10^3$seconds. In this case, the noise decreases by the factor
$\sqrt{\delta \omega\delta t_{ob}/2\pi}\simeq 4.9\times 10^3$, yielding

\begin{equation} 
\frac{I\big(\sigma=(10^{-3}\text{eV}/\sqrt{3})\big)
\sqrt{\frac{\delta \omega\delta t_{ob}}{2\pi}}}{I_n\big(\sigma=(10^{-3}\text{eV}/\sqrt{3})\big)}\simeq 2.5.
\end{equation}

\vspace{0.1cm}

Achieving an electrical conductivity of order $10^{-3}$eV
in semiconductors at $20$mK is challenging, but becomes feasible at $4$K. Although thermal noise is higher at this temperature, the signal can be correspondingly enhanced when we take large radius $R$.
Since the induced current scales with the square of the sample radius, the use of larger samples, i.e. $R=50$cm, significantly increases the signal. Consequently, a large signal-to-noise ratio can be achieved even at $4$K.

\begin{eqnarray}
\frac{I\big(\sigma=\big(10^{-3}\text{eV}/\sqrt{3})\big)}{I_n\big({\sigma=(10^{-3}\text{eV}/\sqrt{3})\big)}}&\times &
\sqrt{\frac{\delta \omega \delta t_{ob}}{2\pi}} \simeq  
0.98g_{\gamma}\Big(\frac{4\text{K}}{T}\Big)^{1/2}\Big(\frac{L}{100\text{cm}}\Big)^{1/2}\Big(\frac{R}{50\text{cm}}\Big)\nonumber \\
&\times &\Big(\frac{B_0}{10\mbox{T}}\Big) \Big(\frac{2\big(\frac{\epsilon}{10}\big)
\big(\frac{\sqrt{3}\sigma}{10^{-3}\text{eV}}\big)^{1/2}\big(\frac{m_a}{10^{-4}\text{eV}}\big)^2}{\big(\frac{\epsilon}{10}\big)^2\big(\frac{m_a}{10^{-4}\text{eV}}\big)^2
+\big(\frac{\sqrt{3}\sigma}{10^{-3}\text{eV}}\big)^2}\Big)
\Big(\frac{\rho_a}{0.3\rm GeVcm^{-3}}\Big)^{1/2}
\Big(\frac{\delta t_{ob}}{1000\,\text{s}}\Big)^{1/2}.	
\end{eqnarray}

\vspace{0.1cm}

 Our method, which employs a cylinder with low electrical conductivity, is also applicable to the detection 
of axion like particle ( ALP ) or so-called dark photon. The electromagnetic coupling of theses particles produces a signal similar to that of the axion.

In the next section, we briefly explain the property of QDC axion coupled with electromagnetic fields.
Using the coupling, we solve electromagnetic fields generated inside the cylinder under strong magnetic field and calculate 
electric power $P$ and current $I$ in section (\ref{s3}). In the section (\ref{s4}), we discuss   
realistic choices of materials forming the large cylinder $R=80$cm and temperature $4$K cooling the cylinder.
We present a figure in which detectable regions of ALPs are shown in $g_{a\gamma\gamma}- m_a$ plane. 
We also show that ALPs with extremely small mass, e.g. $10^{-14}$eV can be detectable with the choice of
electric conductivity being of the order of the mass. 

\section{axion}
The main difficulty in detecting axion dark matter lies in its extremely weak coupling to electromagnetic radiation or ordinary matter ( electrons or nucleons ). In particular, the interaction between the axion field $a(t,\vec{x})$ and the electromagnetic field is described by

\begin{equation}
\label{La}
L_{a\gamma\gamma} = g_{a\gamma\gamma} a(t,\vec{x}) \vec{E}\cdot\vec{B},
\end{equation}
where $\vec{E}$ and $\vec{B}$ denote the electric and magnetic fields, respectively. The coupling constant is
$g_{a\gamma\gamma} = g_{\gamma} \alpha/\pi f_a$,
with the fine structure constant $\alpha \simeq 1/137$, the axion decay constant $f_a$, 
and the relation $m_a f_a \simeq 6 \times 10^{-3}\text{GeV}^2$ for the QCD axion model. 
The model dependent coefficient is $g_\gamma \simeq 0.37$ for the DFSZ model \cite{dfsz,dfsz1} and $g_\gamma \simeq -0.96$ for the KSVZ model \cite{ksvz,ksvz1}.

For a classical axion field representing dark matter, the interaction term $g_{a\gamma\gamma}a(t,\vec{x})$ is extremely small. Assuming that dark matter consists entirely of axions, the local energy density of the dark matter axion is
\begin{equation}
\rho_d = m_a^2 \overline{a(t,\vec{x})^2} = \frac{1}{2} m_a^2 a_0^2 \simeq 0.3\text{GeV}/\text{cm}^3,
\end{equation}
where the overline denotes time averaging. This yields an effective CP-violating interaction of order
$g_{a\gamma\gamma} a(t,\vec{x}) \sim 10^{-21}$,
essentially independent of the QCD axion mass. Consequently, the axion-induced electric field in vacuum under an external magnetic field $B_0$ is extremely weak, of order $\sim g_{a\gamma\gamma} a B_0$.

%
%

\section{electric power generated in cylinder with small conductivity}
\label{s3}
In this letter we show that, for a cylindrical sample in Fig.\ref{cylinder} with small electrical conductivity, the electric power induced by axion dark matter can be large and may be detectable at low temperatures $T \sim 10\text{mK}$. The current flows throughout the bulk of the cylinder, while for a good conductor with large conductivity, 
the current is confined to a thin surface layer.

We focus on the axion mass range $m_a= 10^{-4}\text{-} 10^{-3}\text{eV}$, using a cylinder of length $L = 100\text{cm}$, 
radius $R=6$ cm, and conductivity $\sigma = 10 m_a =10^{-3}\text{-} 10^{-2}\text{eV}$.
As we show soon later, 
the condition of $\sigma=\epsilon m_a$
maximize the electric power induced in the cylinder with permittivity $\epsilon$. 
Such a matter with small conductivity must be realized at low temperatures 
$\sim 10\text{mK}$, for instance by using semiconductors with impurity doping appropriately.

\vspace{0.1cm}
A strong external magnetic field $\vec{B}_0$ is applied parallel to the cylinder axis, so that the system is axially symmetric.
$\vec{B}_0 = (0,0,B_0)$ in cylindrical coordinates $(\rho,\theta,z)$, with $\rho = 0$ at the center and $\rho = R$ at the surface of the cylinder.
In the presence of dark matter axion, the magnetic field induces an oscillating current parallel to $\vec{B}_0$. This current produces microwave radiation with frequencies corresponding to axion masses in the range $m_a =10^{-4}\text{-}10^{-3}\text{eV}$.

We calculate the axion-induced electric field $E'$ to obtain oscillating electric current.

\begin{figure}[htp]
\centering
\includegraphics[width=0.65\hsize]{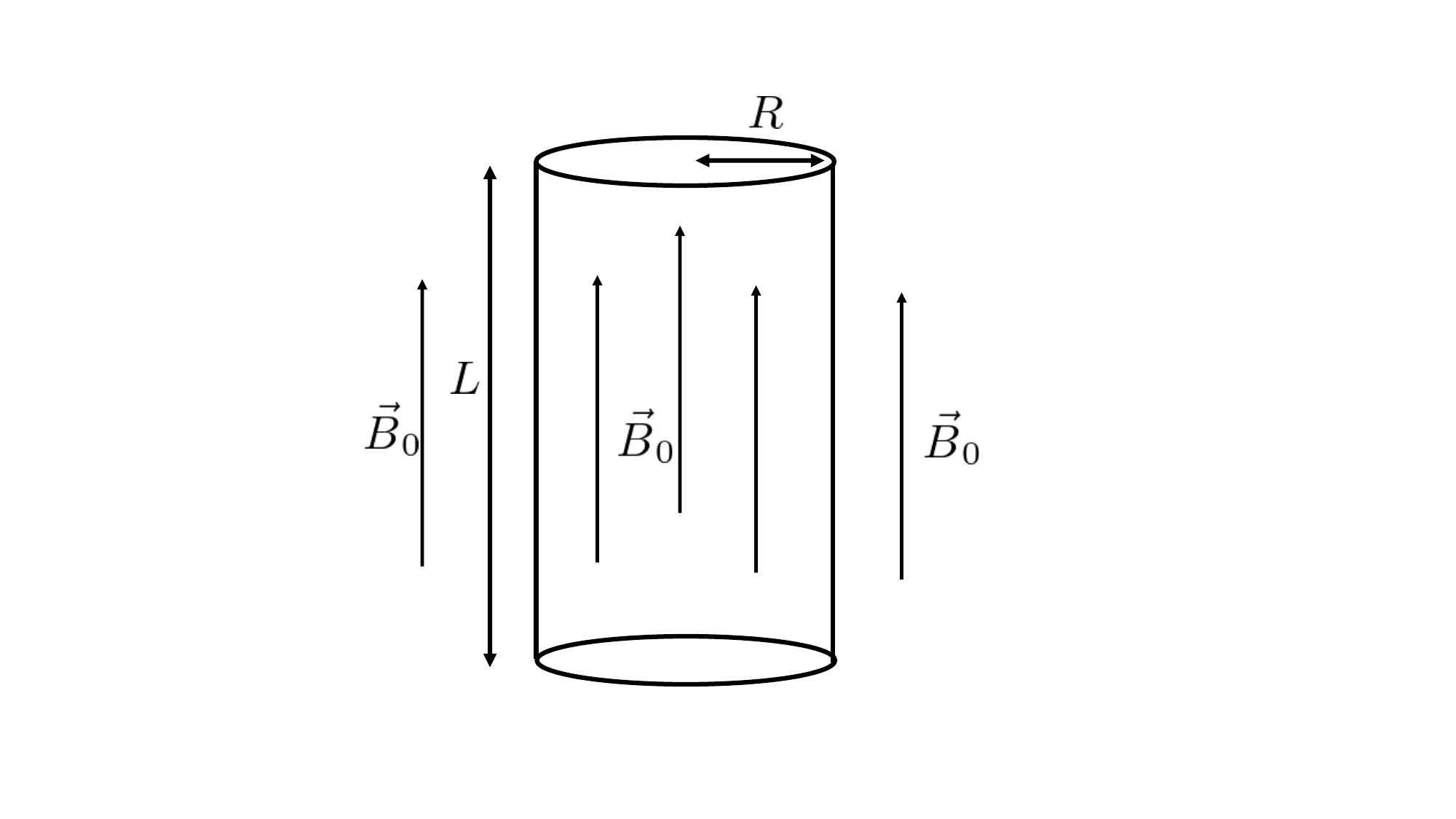}
\caption{cylinder sample with length $L$ and radius $R$ under external magnetic field $\vec{B}_0$}
\label{cylinder}
\end{figure}

\vspace{0.1cm}
 
\vspace{0.1cm}
In order to obtain oscillating electric current $\vec{J}=\sigma \vec{E}'$, we solve the Maxwell equations involving axion effect,
 
\begin{eqnarray}
\label{modified}
\vec{\partial}\cdot(\epsilon\vec{E}'+g_{a\gamma\gamma}a(t,\vec{x})\vec{B})&=0&, \quad 
\vec{\partial}\times \Big(\vec{B}-g_{a\gamma\gamma}a(t,\vec{x})\vec{E}'\Big)-
\partial_t\Big(\epsilon\vec{E}'+g_{a\gamma\gamma}a(t,\vec{x})\vec{B}\Big)=\vec{J},   \\
\vec{\partial}\cdot\vec{B}&=0&, \quad \vec{\partial}\times \vec{E}'+\partial_t \vec{B}=0
\end{eqnarray}
where permittivity $\epsilon$ and electric current $\vec{J}$ are non vanishing inside the cylinder,
while $\epsilon=1$ and $\vec{J}=0$ in vacuum. We have assumed trivial permeability $\mu=1$ of the cylinder. 

The electric field $\vec{E}'$ induced by the axion effect is much small, being of the order of $g_{a\gamma\gamma}aB_0$,
while $\vec{B}=\vec{B}_0+\vec{B}'$ with external magnetic field $\vec{B}_0=(0,0,B_0)$. 
$\vec{B}'$ is induced by the axion and associated with $\vec{E}'$. 

It is easy to obtain the following equation of electric field $\vec{E}'$ using cylindrical coordinate.

\begin{equation}
\big(\partial_{\rho}^2+\frac{1}{\rho}\partial_{\rho}+\epsilon m_a^2+i\sigma m_a\big)\vec{E}'=-m_a^2g_{a\gamma\gamma}a(t)\vec{B}_0
\end{equation}
assuming $a(t)\propto \exp(-im_at)$. The solution $\vec{E}'=(0,0,E')$ is

\begin{equation}
E'=d(t)J_0(bm_a\rho)+\frac{m_a^2E_a}{\epsilon m_a^2+im_a\sigma}
\end{equation}
with $E_a\equiv -g_{a\gamma\gamma}a(t)B_0$,
where $d(t)\propto \exp(-im_at)$ is a constant determined by boundary conditions at $\rho=R$.
The corresponding magnetic field $\vec{B}'=(0,B',0)$ is given by solving $\partial_t B'=\partial_{\rho}E'$, 

\begin{equation}
B'=-ib\,d(t)J_1(bm_a\rho)
\end{equation}

In the above expression, $J_{0,1}(x)$ denotes Bessel function of the first kind and is chosen 
because of its finiteness of $E'(\rho)$ at $\rho=0$.
The constant $b$ is given by

\begin{equation}
b\equiv (\epsilon^2+y^2)^{1/4}\exp(i\theta/2) \quad \mbox{with} \quad \theta=\cos^{-1}\Big(\frac{\epsilon}{\sqrt{\epsilon^2+y^2}}\Big)
\end{equation}
with $y\equiv \sigma/m_a$.

These solutions $E'$ and $B'$ represent electric and magnetic fields inside the cylinder with radius $R$.
Similarly, we can find solutions $E_v$ and $B_v$ outside the cylinder with the conditions $\vec{J}=0$ and $\epsilon =1$ in the
Maxwell equations.

\begin{equation}
E_v=\tilde{d}(t)H_0^{(1)}(m_a\rho)+E_a \quad \mbox{and} \quad B_v=-i\tilde{d}(t)H_1^{(1)}(m_a\rho)
\end{equation}
with $H_{0,1}^{(1)}$ Hankel functions of the first kind,
where $\tilde{d}\propto \exp(-im_at)$ is a constant determined by boundary conditions at $\rho=R$.
The Hankel function of the first kind is chosen because the radiations described by $E_v$ and $B_v$ are
outgoing waves; $E_v( B_v) \sim \exp(-im_at+im_a\rho)$ as $\rho \to \infty$.

\vspace{0.1cm}
In order to determine the constants $d(t)$ and $\tilde{d}(t)$, we impose boundary conditions at the surface $\rho=R$
of the cylinder such that $\epsilon E'(\rho=R)=E_v(\rho=R)$ and $B'(\rho=R)=B_v(\rho=R)$.
Then, we have

\begin{equation}
\epsilon d(t)J_0(bm_aR)+\frac{\epsilon m_a^2E_a}{\epsilon m_a^2+im_a\sigma}=\tilde{d}(t)H_0^{(1)}(m_aR)+E_a \quad
\mbox{and} \quad  b\,d(t)J_1(bm_aR)=\tilde{d}(t)H_1^{(1)}(m_aR)
\end{equation}

Therefore, by solving the equations for $d(t)$ and $\tilde{d}(t)$,
we have

\begin{equation}
d(t)=\frac{iy}{\epsilon+iy}\frac{E_aH_1^{(1)}(x)}{\epsilon J_0(bx)H_1^{(1)}(x)-bJ_1(bx)H_0^{(1)}(x)}, \quad
\tilde{d}(t)=d(t)\Bigg(\frac{bJ_1(bx)}{H_1^{(1)}(x)}\Bigg),
\end{equation}
with $x=m_aR$.

Thus, 
the oscillating electric and magnetic fields $E'$ and $B'$ inside the cylinder are

\begin{eqnarray}
\label{13}
E'(\rho)&=&E_a\Big(\frac{1}{\epsilon+iy}+\frac{iy}{\epsilon+iy}\frac{H_1^{(1)}(m_aR)J_0(bm_a\rho)}{\epsilon J_0(bm_aR)H_1^{(1)}(m_aR)
-bJ_1(bm_aR)H_0^{(1)}(m_aR)}\Big)  \\
B'(\rho)&=&E_a\Big(\frac{by}{\epsilon+iy}\frac{E_aH_1^{(1)}(m_aR)J_1(bm_a\rho)}{\epsilon J_0(bm_aR)H_1^{(1)}(m_aR)
-bJ_1(bm_aR)H_0^{(1)}(m_aR)}\Big).
\end{eqnarray}
They oscillate as $E'(B') \propto \exp(-i m_a t)$. Obviously, the first term of the electric field, $E'(\rho)$, exists throughout the bulk of the cylinder, whereas the second term is confined to its surface; 
$J_0(bm_a\rho)/J_{0,1}(bm_aR)\propto \exp\big((R-\rho)m_a\text{Im}(b)\big)$ 
as $\rho \to \infty$ where $\text{Im}(O)$ denotes imaginary part of $O$.

\vspace{0.1cm}
Uniquely, the axion–electromagnetic coupling drives an oscillating electric field that permeates the entire cylindrical sample, in contrast to ordinary electromagnetic induction, where the high frequency electric field is confined to the surface.
A similar fact is also present in the case of dark photon\cite{kishimoto}.

As shown later, the second term dominates the electric power ( Joule heat ) in the limit of large conductivity $(\sigma \sim 10^4\text{eV})$, while the first term becomes dominant for small conductivity $(\sigma \sim 10^{-3}\text{eV})$.

\vspace{0.1cm}
Using the formula of the electric field $E'$, we obtain oscillating electric current $J(\rho)=\sigma Re (E'(\rho))$ and corresponding
its power $P$ averaged over the period $2\pi/m_a$,

\begin{equation}
P=\int_0^{R}\overline{J(\rho)(Re (E'(\rho)))}L 2\pi \rho d\rho=\int_0^{R}\sigma |E'(\rho)|^2L \pi \rho d\rho
\end{equation}
where $Re(O)$ denotes real part of the quantity $O$.

When we put $z=\rho/R$, the formula $P$ is rewritten such that

\begin{equation}
P=\sigma L\pi \int_0^{1}|E'(Rz)|^2R^2 z dz=\frac{\pi L|E_a|^2}{m_a}\int_0^{1}yx^2|U(x,y,z)|^2zdz ,
\end{equation}
where $U(x,y,z)$ is

\begin{equation}
\label{17}
U(x,y,z)\equiv \frac{1}{\epsilon+iy}+\frac{iy}{\epsilon+iy}\frac{H_1^{(1)}(x)J_0(bxz)}{\epsilon J_0(bx)H_1^{(1)}(x)
-bJ_1(bx)H_0^{(1)}(x)} 
\end{equation}
We remind that $y\equiv \sigma/m_a$ and $x\equiv m_aR$ and $b=b(y)$ is the function of $y$.
$P$ is a complicated function in $x$ and $y$, or $m_a$ and $R$. But we show below that $P$ becomes a simple function
in $x$ and $y$ when we take the limit of large conductivity,
$\sigma\sim 10^4$eV or small one $\sigma\sim 10^{-3}$eV. We note that electric field is written such that $E'(\rho)=E_aU$.

It is easy to confirm that in the limit of $\sigma \to \infty$ ( $y\to \infty$ ), 
$P\propto m_a\rho_a V(B_0/m_aM)^2|H_1^{(1)}(x)/H_0^{(1)}(x)|^2$ with $M\equiv \pi f_a/(g_{\gamma}\alpha)=1/g_{a\gamma\gamma}$ 
and the volume $V=2\pi RL\delta$,
within which the electric current flows. $\delta=\sqrt{2/m_a\sigma}$ denotes skin depth. 
The current flows only in the surface with the depth $\delta$ of the cylinder. It is coincident with the previous result \cite{kishimoto}.
In the limit, the main contribution comes from the second term in $U$. We note that $|H_1^{(1)}(x)/H_0^{(1)}(x)|^2\simeq 1$ for
$x>5$ and that $b(y)\to \sqrt{\sigma/m_a}\exp(i\pi/4)$ in the limit $y=\sigma/m_a \gg 1$.

\vspace{0.1cm}
The skin depth for general $\sigma$ represented by the second term of $U(x,y,z)$ in eq(\ref{17})
is given by 
$\big(m_a \mbox{Im}(b(y))\big)^{-1}=\Big(m_a(\epsilon^2+(\sigma/m_a)^2)^{1/4}\sin(\theta/2)\Big)^{-1}>\delta=\sqrt{2/m_a\sigma}$.
In this paper, assuming typical value $\epsilon=10$ in semiconductor, 
we take small electric conductivity $\sigma \sim \epsilon m_a\sim 10m_a$ ( $y\sim 10$ ) and 
consider the mass region $m_a=10^{-3}\mbox{-}10^{-4}\mbox{eV}$. Then, the skin depth 
$\big(m_a \mbox{Im}(b(y))\big)^{-1}\simeq 0.1\text{cm}(10^{-4}\text{eV}/m_a)$, smaller than the radius $R=6$cm examined in our paper.
On the other hand, the first term of $U$ is present in the whole of the cylinder, not confined at the surface.

\vspace{0.1cm}
In such a cylinder with small electrical conductivity,
we find that the main contribution to $P$ comes from the first term in $U$.
Actually, the following quantities are monotonically increasing function in $x$,

\begin{eqnarray}
&\int_0^1& yx^2dz z\Bigg|\frac{iy}{\epsilon+iy}\frac{H_1^{(1)}(x)J_0(bxz)}{\epsilon J_0(bx)H_1^{(1)}(x)
-bJ_1(bx)H_0^{(1)}(x)}\Bigg|^2 =0.01\text{-} 0.09 \,\,\, \mbox{for} \,\,\, x=1\text{-} 10 \\
&\int_0^1& yx^2dz z\Big|\frac{1}{\epsilon+iy}\Big|^2=0.025\text{-} 2.5 \,\,\, \mbox{for} \,\,\, x=1\text{-} 10.
\end{eqnarray}
with $y=10$ and $\epsilon=10$. We find that for each $x$, the quantity in the second equation is larger than the one in 
the first equation. Especially, it is more than an order of magnitude for $x \ge10$. Further,
smaller $y$ ( $<10$ ) leads to much larger discrepancy between the quantities.
Therefore, we may write

\begin{equation}
\label{20}
P=\frac{\pi L|E_a|^2}{m_a}\int_0^{1}yx^2|U(x,y,z)|^2zdz\sim \frac{\pi L|E_a|^2}{2m_a}\Big(\frac{yx^2}{\epsilon^2+y^2}\Big).
\end{equation}
 
$P$ takes the maximal value at $y=\sigma/m_a=\epsilon$ 
and $P(\epsilon,y=\epsilon)\propto \epsilon^{-1}$. Thus, it is favorable to take the conductivity $\sigma =\epsilon m_a$.
Because we do not know the value of the axion mass $m_a$, the value $\sigma=\epsilon m_a$ is unknown. But when we put
$\sigma=\epsilon m_a$ in the formula, the dependence of $P$ on $m_a$ becomes simple; $P\propto m_a$. 
Otherwise, the dependence of $P$ on $m_a$ or $\sigma$ is not simply scaling; $P\propto \sigma m_a^2/(\epsilon^2m_a^2+\sigma^2)$.
In the actual search of the axion mass
in the range $10^{-4}\text{-} 10^{-3}$eV, it is sufficient for
the axion detection to take the value $\sigma\sim 1.0\times 10^{-3}$eV with $\epsilon \sim10$. 
Although it does not lead to the maximal power $P$,
the predicted signal-noise ratios are large enough for the detection. 
Hereafter, we take $\sigma=\epsilon m_a$ and $\epsilon=10$ to estimate the power $P$.
Later we find  that  $\sigma=\epsilon m_a/\sqrt{3}$ maximizes the ratio of electric current $I$ induced by axion to thermal
noise $I_n$. Please refer Fig.2 in which the dependence of $P(\sigma, m_a)$ and the current $I(\sigma, m_a)$ on $\sigma$ and $m_a$
is shown. 
\vspace{0.1cm}

Numerically, noting $|E_a|=|g_{a\gamma\gamma}a(t)B_0|\simeq 2.4\times 10^{-36}\text{GeV}^2g_{\gamma}(B_0/15\text{T})$, we find 

\begin{equation}
\label{21}
P(\sigma=10^{-3}\text{eV})\simeq 5.3\times 10^{-27}\mbox{W}g_{\gamma}^2
\Big(\frac{L}{100\mbox{cm}}\Big)\Big(\frac{R}{6\mbox{cm}}\Big)^2
\Big(\frac{B_0}{15\mbox{T}}\Big)^2
\Big(\frac{2\big(\frac{\sigma}{10^{-3}\text{eV}}\big)\big(\frac{m_a}{10^{-4}\text{eV}}\big)^2}{\big(\frac{\epsilon}{10}\big)^2\big(\frac{m_a}{10^{-4}\text{eV}}\big)^2+\big(\frac{\sigma}{10^{-3}\text{eV}}\big)^2}\Big)\Big(\frac{\rho_a}{0.3\rm GeVcm^{-3}}\Big)
\end{equation}
with $y=10$ ( $\sigma=10m_a$ ) and $x\simeq 30$ ( $x=m_aR=10^{-3}\text{eV}\times 6\text{cm}\simeq 30$ ).
Obviously, $P(\sigma=10^{-3}\text{eV})$ in eq.(\ref{21}) is proportional to $R^2=x^2/m_a^2$,
that is, the cross section $\pi R^2$ of the cylinder. The energy dissipation takes place in the bulk volume $\propto R^2L$, not merely in the surface, resulting in substantially enhanced dissipation power $P$.

\vspace{0.1cm}
For comparison, we present the power $P$
when the conductivity is much high, $\sigma =10^4$eV,

\begin{eqnarray}
\label{22}
P(\sigma=10^4\text{eV})&\simeq& |E_a|^2\frac{\pi m_a \delta R L}{2}\Bigg|\frac{H_1^{(1)}(x)}{H_0^{(1)}(x)}\Bigg|^2 \nonumber \\
&\simeq& 5.4\times 10^{-31}\mbox{W}g_{\gamma}^2\Big(\frac{L}{100\mbox{cm}}\Big)\Big(\frac{R}{6\mbox{cm}}\Big)
\sqrt{\frac{m_a}{10^{-4}\mbox{eV}}}\sqrt{\frac{10^4\mbox{eV}}{\sigma}}
\Big(\frac{B_0}{15\mbox{T}}\Big)^2\Big(\frac{\rho_a}{0.3\rm GeVcm^{-3}}\Big)
\end{eqnarray} 
with $y=\sigma /m_a=10^8$ and $x=m_aR\simeq 30$.
It is generated in the surface with the skin depth $\delta\sim 10^{-5}$cm of the cylinder.
We find that the power $P(\sigma=10^{-3}\text{eV})$
is fourth order of magnitude larger than $P(\sigma=10^{4}\text{eV})$.
That is, $P(\sigma=10^{-3}\text{eV})>10^4P(\sigma=10^4\text{eV})$. 
It is the reason why we use such a cylinder with small conductivity $\sigma\sim 10^{-3}$eV.

\vspace{0.1cm}
There are two reasons of the small power  $P(\sigma=10^4\text{eV})$ in the cylinder. One is the suppression of electric
field inside conductor with large conductivity $\sigma=10^4$eV, that is, $\sqrt{m_a/\sigma}|E_a|$ \cite{aiwazaki,kishimoto}
compared with $|E_a|/2\epsilon$ in material with small conductivity $\sigma=10^{-3}$eV. The other one is
the presence of the skin depth $\delta=\sqrt{2/m_a\sigma}$. The area $2\pi R \delta$ of the current flowing is 
much smaller than that $\pi R^2$ in material with low conductivity.
Therefore, even if large conductivity $\sigma$ amplifies electric current by the factor $\sigma$, both the suppression of
electric field by the factor $\sqrt{m_a/\sigma}$ and the area of current flow by the factor $\sqrt{1/m_a\sigma}$, cause
the power $P(\sigma=10^4\text{eV})$ smaller than the power $P(\sigma=10^{-3}\text{eV})$. It is roughly,

\begin{equation}
\frac{P(\sigma_h=10^4\text{eV})}{P(\sigma=10^{-3}\text{eV})}\sim \frac{ R\delta \sigma_h (\sqrt{m_a/\sigma_h})^2 }
{R^2\sigma (1/\epsilon^2)}= \frac{\delta m_a\epsilon^2 }{R\sigma}\sim 10^{-3} 
\end{equation}
with $R=1$cm, $\epsilon=10$ and $m_a=10^{-4}$eV. It is coincident with the above calculations in eq(\ref{21}) and eq(\ref{22}).

\vspace{0.1cm}
The large power $P(\sigma=10^{-3}\text{eV})$ arises mainly from large electric current $I$ flowing in the bulk of the cylinder.
The electric current $I$ is,
\begin{equation}
I(\sigma=10^{-3}\text{eV})=\frac{\sigma\epsilon |E_a|S}{(\epsilon^2+y^2)}\simeq 2.9\times 10^{-14}\text{A}g_{\gamma}\Big(\frac{R}{6\text{cm} }\Big)^2\Big(\frac{2\big(\frac{\epsilon}{10}\big)\big(\frac{\sigma}{10^{-3}\text{eV}}\big)\big(\frac{m_a}{10^{-4}\text{eV}}\big)^2}{\big(\frac{\epsilon}{10}\big)^2\big(\frac{m_a}{10^{-4}\text{eV}}\big)^2+\big(\frac{\sigma}{10^{-3}\text{eV}}\big)^2}\Big)
\Big(\frac{B_0}{15\text{T}}\Big)\Big(\frac{\rho_a}{0.3\rm GeVcm^{-3}}\Big)^{1/2},
\end{equation}
where the electric field $E=Re(E')=\epsilon|E_a|/(\epsilon^2+y^2)=|E_a|/20$ is the one in the first term with $y=10$ in eq(\ref{13}). 

\vspace{0.1cm}
The point causing the large current $I(\sigma=10^{-3}\text{eV})$ is that electric field $E$ inside of the cylinder with small conductivity, e.g. $\sigma=10^{-3}$eV
has no suppression factor $\sqrt{m_a/\sigma}\simeq 10^{-4}$, which is present when the cylinder with large conductivity $\sigma=10^4$eV.
Additionally, the current flows in the bulk of the cylinder, not just in the surface.
The facts make the large electric current $I(\sigma=10^{-3}\text{eV})\sim 10^{-14}$A.

\vspace{0.1cm}

Figure 2 shows the dependence of the power $P(\sigma, m_a)$ and current $I(\sigma, m_a)$ 
on the axion mass $m_a$ and electrical conductivity $\sigma$ when $\epsilon=10$. The vertical axis is given in arbitrary units. 
It is remarkable feature that
as the axion mass $m_a$ increases, increasing the electrical conductivity $\sigma$ correspondingly leads to 
larger values of both $P(\sigma, m_a)$ and $I(\sigma, m_a)$.


\begin{figure}[htp]
\label{2f}
\centering
\includegraphics[width=0.60\hsize]{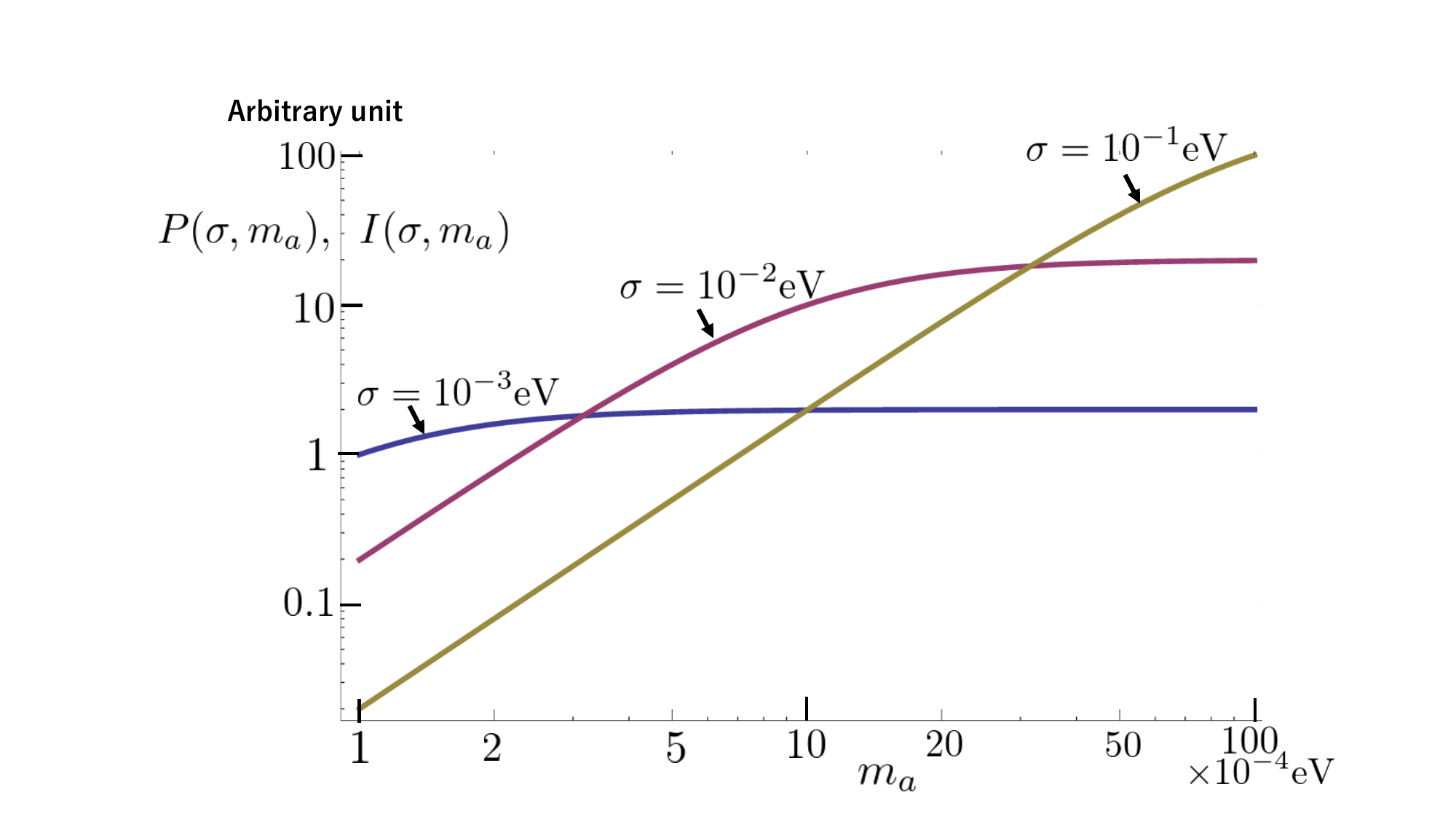}
\caption{Dependence of power $P(\sigma, m_a)$ and current $I(\sigma, m_a)$ on $\sigma$ and $m_a$ when $\epsilon=10$. }
\label{cylinder}
\end{figure}

\vspace{0.1cm}
In order to observe the power $P(\sigma=10^{-3}\text{eV})$, we need to take into account thermal noise  ( Johnson-Nyquist noise ).
The noise power $P_n$ is given such that $P_n=T\delta \omega/2\pi$ with $\delta \omega=10^{-6}m_a$.
$\delta \omega$ denotes frequency width in the observation.

Then we have

\begin{equation}
P_n=\frac{T\delta \omega}{2\pi} \simeq 6.6\times 10^{-21}\text{W}\Big(\frac{T}{20\text{mK}}\Big)
\Big(\frac{m_a}{10^{-4}\text{eV}}\Big).
\end{equation}

The noise is reduced by the factor $\sqrt{\delta \omega \delta t_{ob}/2\pi}$ with
observation time $\delta t_{ob}$. It yields

\begin{equation}
P_n \to \frac{P_n}{\sqrt{\frac{\delta \omega \delta t_{ob}}{2\pi}}}\simeq 4.2\times 10^{-24}\text{W}
\Big(\frac{T}{20\text{mK}}\Big)\Big(\frac{m_a}{10^{-4}\text{eV}}\Big)^{1/2}\Big(\frac{100\,\text{s}}{\delta t_{ob}}\Big)^{-1/2}.
\end{equation}

Therefore, the ratio of $P(\sigma=10^{-3}\text{eV})$ to the noise $P_n/\sqrt{\delta \omega \delta t_{ob}/2\pi}$ is

\begin{eqnarray}
&&\frac{P(\sigma=10^{-3}\text{eV})}{P_n}\sqrt{\frac{\delta \omega \delta t_{ob}}{2\pi}}  \simeq 1.3\times 10^{-3}g_{\gamma}^2
\Big(\frac{20\text{mK}}{T}\Big)\Big(\frac{L}{100\mbox{cm}}\Big)\Big(\frac{R}{6\mbox{cm}}\Big)^2\times  \nonumber \\
&& \Big(\frac{B_0}{15\mbox{T}}\Big)^2\Big(\frac{2\big(\frac{\sigma}{10^{-3}\text{eV}}\big)\big(\frac{m_a}{10^{-4}\text{eV}}\big)^{3/2}}{\big(\frac{\epsilon}{10}\big)^2\big(\frac{m_a}{10^{-4}\text{eV}}\big)^2+\big(\frac{\sigma}{10^{-3}\text{eV}}\big)^2}\Big)
\Big(\frac{\rho_a}{0.3\rm GeVcm^{-3}}\Big)\Big(\frac{\delta t_{ob}}{100\,\text{s}}\Big)^{1/2}.
\end{eqnarray}

It is the ratio when $m_a=10^{-4}$eV and $\sigma=10^{-3}$eV. Further, when $m_a=10^{-3}$eV, taking $\sigma=10m_a=10^{-2}$eV, 
we have 

\begin{eqnarray}
\label{31}
&&\frac{P(\sigma=10^{-2}\text{eV})}{P_n}\sqrt{\frac{\delta \omega \delta t_{ob}}{2\pi}} \simeq 4.1\times 10^{-3}g_{\gamma}^2
\Big(\frac{20\text{mK}}{T}\Big)\Big(\frac{L}{100\mbox{cm}}\Big)\Big(\frac{R}{6\mbox{cm}}\Big)^2\times  \nonumber \\
&& \Big(\frac{B_0}{15\mbox{T}}\Big)^2\Big(\frac{2\big(\frac{\sigma}{10^{-2}\text{eV}}\big)\big(\frac{m_a}{10^{-3}\text{eV}}\big)^{3/2}}{\big(\frac{\epsilon}{10}\big)^2\big(\frac{m_a}{10^{-3}\text{eV}}\big)^2+\big(\frac{\sigma}{10^{-2}\text{eV}}\big)^2}\Big)
\Big(\frac{\rho_a}{0.3\rm GeVcm^{-3}}\Big)\Big(\frac{\delta t_{ob}}{100\,\text{s}}\Big)^{1/2}.
\end{eqnarray}

\vspace{0.1cm} 

SN ratio of the power is less than $10^{-2}$, but we can show that SN ratio of electric current $I$ is bigger than $1$.

\vspace{0.1cm}
We should make a comment that we have taken $y\equiv \sigma/m_a=\epsilon$ in order to maximize the SN ratio $P/P_n$.
On the other hand, to maximize SN ratio $I/I_n$ of the electric current to thermal noise $I_n$, we need to take $y=\epsilon/\sqrt{3}$,
because $I/I_n\propto yy^{-1/2}/(\epsilon^2+y^2)$:
$I_n=\sqrt{2T\delta \omega/\pi R_c}$ with the resistance of the cylinder, $R_c=L/(\sigma \pi R^2)$.
However, we take $y=\epsilon$ for simplicity in the subsequent calculations. It simply causes slightly smaller value of the
ratio $I/I_n$; $y^{1/2}/(\epsilon^2+y^2)\simeq 1.58\times 10^{-3}$ with $y=\epsilon=10$, 
while  $y^{1/2}/(\epsilon^2+y^2)\simeq 1.8\times 10^{-3}$ with $y=\epsilon/\sqrt{3}$.

\vspace{0.1cm}

Because thermal noise $I_n$ is 

\begin{equation}
 I_n(\sigma=10^{-3}\text{eV})=\sqrt{\frac{2T\delta \omega}{\pi R_c}}\simeq 6.9\times 10^{-11}\text{A}\Big(\frac{T}{20\text{mK}}\Big)^{1/2}
\Big(\frac{100\text{cm}}{L}\Big)^{1/2}
\Big(\frac{R}{6\text{cm}}\Big) \Big(\frac{\sigma}{10^{-3}\text{eV}}\Big)^{1/2}\Big(\frac{m_a}{10^{-4}\text{eV}}\Big)^{1/2}    
\end{equation}
with the resistance $R_c\simeq 1.44(L/100\text{cm})(6\text{cm}/R)^2(10^{-3}\text{eV}/\sigma)\simeq 540\Omega$, 
the SN ratio is  

\begin{eqnarray}
\frac{I(\sigma=10^{-3}\text{eV})}{I_n({\sigma=10^{-3}\text{eV})}}&\sqrt{\frac{\delta \omega \delta t_{ob}}{2\pi}}& 
\simeq  
2.1g_{\gamma}\Big(\frac{20\text{mK}}{T}\Big)^{1/2}\Big(\frac{L}{100\text{cm}}\Big)^{1/2}\Big(\frac{R}{6\text{cm}}\Big)\nonumber \\
&\times \Big(\frac{B_0}{15\text{T}}\Big)& \Big(\frac{2\big(\frac{\epsilon}{10}\big)\big(\frac{\sigma}{10^{-3}\text{eV}}\big)^{1/2}\big(\frac{m_a}{10^{-4}\text{eV}}\big)^2}{\big(\frac{\epsilon}{10}\big)^2\big(\frac{m_a}{10^{-4}\text{eV}}\big)^2+\big(\frac{\sigma}{10^{-3}\text{eV}}\big)^2}\Big)
\Big(\frac{\rho_a}{0.3\rm GeVcm^{-3}}\Big)^{1/2}
\Big(\frac{\delta t_{ob}}{1000\,\text{s}}\Big)^{1/2}	
\end{eqnarray}
It is the ratio corresponding to axion mass $m_a=10^{-4}$eV, while it is given by 

\begin{eqnarray}
\frac{I(\sigma=10^{-2}\text{eV})}{I_n({\sigma=10^{-2}\text{eV})}}\times &\sqrt{\frac{\delta \omega \delta t_{ob}}{2\pi}}& 
\simeq  
6.6g_{\gamma}\Big(\frac{20\text{mK}}{T}\Big)^{1/2}\Big(\frac{L}{100\text{cm}}\Big)^{1/2}\Big(\frac{R}{6\text{cm}}\Big)\nonumber \\
&\times &\Big(\frac{B_0}{15\mbox{T}}\Big)\Big(\frac{2\big(\frac{\epsilon}{10}\big)\big(\frac{\sigma}{10^{-2}\text{eV}}\big)^{1/2}\big(\frac{m_a}{10^{-3}\text{eV}}\big)^2}{\big(\frac{\epsilon}{10}\big)^2\big(\frac{m_a}{10^{-3}\text{eV}}\big)^2+\big(\frac{\sigma}{10^{-2}\text{eV}}\big)^2}\Big) 
\Big(\frac{\rho_a}{0.3\rm GeVcm^{-3}}\Big)^{1/2}
\Big(\frac{\delta t_{ob}}{1000\,\text{s}}\Big)^{1/2}	
\end{eqnarray}
for $m_a=10^{-3}$eV.

\vspace{0.1cm}

Consequently, we find that the axion with mass $m_a =10^{-4}\text{-}10^{-3}\text{eV}$ can be detected
by observing the electric current induced in the semiconductor cylinder. Because their frequencies are large such
as $24\text{-}240$GHz, it may be detected with heterodyne detector. 

\section{discussion}
\label{s4}
Increasing the radius $R$ of the cylindrical sample enhances the induced current $I\propto R^2$ and thus improves detectability. However, in a dilution refrigerator operating at $T=20$ mK, the allowable sample size is severely limited. In contrast, at $T=4$ K, large samples can be readily cooled by direct immersion in liquid He.
By applying a strong magnetic field to such large samples using a superconducting solenoid, the axion-induced current can be significantly enhanced, facilitating detection. Moreover, at $20$ mK, achieving a conductivity of order 
$10^{-3}$eV is challenging for semiconductors and may require alternative materials such as amorphous systems. At $4$ K, however, semiconductors with appropriate doping can readily meet this requirement and serve as suitable samples.
For example, we dope phosphorus into silicon at a critical concentration of approximately
$\sim 3.5\times 10^{18}\text{cm}^{-3}$, beyond which the semiconductor becomes a metal.
Adjacent localized wave functions of electrons doped overlap, creating a new conduction band. It leads to the metal-insulator transition.
Thus, we adjust a carrier mobility $\mu\sim 10\text{cm}^2/\text{Vs}$ and may obtain $\sigma=10^{-3}-10^{-2}\text{eV}$.
We remember that the conductivity $\sigma$ is approximately given by $\sigma=en\mu$; $n$ is carrier density. 
It is easier to perform such an adjustment at $4$K than $20$mK. 
  
\vspace{0.1cm}
Taking into account a detectable signal-to-noise ratio, the following experimental setup can be considered.
That is, large sample scale is $L=100$cm and $R=80$cm, and magnetic field $B=7$T.

\begin{eqnarray}
\label{37}
\frac{I\big(\sigma=(10^{-3}\text{eV}/\sqrt{3})\big)}{I_n({\sigma= \big(10^{-3}\text{eV}/\sqrt{3})\big)}}&\times& 
\sqrt{\frac{\delta \omega \delta t_{ob}}{2\pi}} \simeq  
1.1g_{\gamma}\Big(\frac{4\text{K}}{T}\Big)^{1/2}\Big(\frac{L}{100\text{cm}}\Big)^{1/2}  \nonumber \\ 
&\times& \Big(\frac{R}{80\text{cm}}\Big)
\Big(\frac{B_0}{7\mbox{T}}\Big)\Big(\frac{2\big(\frac{\epsilon}{10}\big)\big(\frac{\sqrt{3}\sigma}{10^{-3}\text{eV}}\big)^{1/2}\big(\frac{m_a}{10^{-4}\text{eV}}\big)^2}{\big(\frac{\epsilon}{10}\big)^2\big(\frac{m_a}{10^{-4}\text{eV}}\big)^2
+\big(\frac{\sqrt{3}\sigma}{10^{-3}\text{eV}}\big)^2}\Big)
\Big(\frac{\rho_a}{0.3\rm GeVcm^{-3}}\Big)^{1/2}
\Big(\frac{\delta t_{ob}}{1000\,\text{s}}\Big)^{1/2}	
\end{eqnarray}
for $m_a=10^{-4}$eV, where we have taken $\sigma=\epsilon m_a/\sqrt{3}$ to maximize the ratio.
The electric current is large such as $I\simeq 2.1\times 10^{-12}$A.

Similarly, 

\begin{eqnarray}
\label{38}
\frac{I\big(\sigma=(10^{-2}\text{eV}/\sqrt{3})\big)}{I_n\big({\sigma=(10^{-2}\text{eV}/\sqrt{3})\big)}}&\times &\sqrt{\frac{\delta \omega \delta t_{ob}}{2\pi}} 
\simeq  3.4g_{\gamma}\Big(\frac{4\text{K}}{T}\Big)^{1/2}\Big(\frac{L}{100\text{cm}}\Big)^{1/2} \nonumber \\
&\times &\Big(\frac{R}{80\text{cm}}\Big)
\Big(\frac{B_0}{7\mbox{T}}\Big) \Big(\frac{2\big(\frac{\epsilon}{10}\big)\big(\frac{\sqrt{3}\sigma}{10^{-2}\text{eV}}\big)^{1/2}\big(\frac{m_a}{10^{-3}\text{eV}}\big)^2}{\big(\frac{\epsilon}{10}\big)^2\big(\frac{m_a}{10^{-3}\text{eV}}\big)^2
+\big(\frac{\sqrt{3}\sigma}{10^{-2}\text{eV}}\big)^2}\Big)
\Big(\frac{\rho_a}{0.3\rm GeVcm^{-3}}\Big)^{1/2}
\Big(\frac{\delta t_{ob}}{1000\,\text{s}}\Big)^{1/2}	
\end{eqnarray}
for $m_a=10^{-3}$eV. The electric current is $I\simeq 2.1\times 10^{-11}$A.

\vspace{0.1cm}
Obviously, we need a large superconducting solenoid to incorporate the large sample with $L=100$cm and $R=80$cm.  
But, the sample volume in eq(\ref{37}) and eq(\ref{38}) is too large to be efficiently cooled and to be incorporated in a superconducting solenoid, one can instead use ten smaller samples, each with a radius of 8 cm and a length of 10 cm, connected in parallel. 
In such a configuration, the use of multiple superconducting solenoids,
possibly on the order of ten, may be required instead of a single magnet.
Each one incorporates a small sample. Because the samples are connected in parallel, total electric current 
is identical to the one generated in a large sample.
Then, a comparable signal-to-noise ratio can be achieved. 
In this way, if a large superconducting solenoid capable of accommodating a sample with 
$L=100$cm and 
$R$=80cm, or alternatively several smaller superconducting solenoids accommodating samples with $L=10$cm and 
$R$=8 cm connected in parallel, can be realized, then the detection of axions with masses in the range
$m_a=10^{-4}-10^{-3}$eV is feasible.

\vspace{0.1cm}

Up to now, we have presented results in QCD axion model. More generally, we show SN ratio of electric current in terms
of $g_{a\gamma\gamma}$ and $m_a$,

\begin{eqnarray}
\frac{I\big(\sigma=(10^{-3}\text{eV}/\sqrt{3})\big)}{I_n({\sigma= \big(10^{-3}\text{eV}/\sqrt{3})\big)}}&\times& 
\sqrt{\frac{\delta \omega \delta t_{ob}}{2\pi}} \simeq  
2.9\times 10^{-2}\Big(\frac{g_{a\gamma\gamma}}{10^{-15}\text{GeV}^{-1}}\Big)
\Big(\frac{4\text{K}}{T}\Big)^{1/2}\Big(\frac{L}{100\text{cm}}\Big)^{1/2}  \nonumber \\ 
&\times& \Big(\frac{R}{80\text{cm}}\Big)
\Big(\frac{B_0}{7\mbox{T}}\Big)\Big(\frac{2\big(\frac{\epsilon}{10}\big)\big(\frac{\sqrt{3}\sigma}{10^{-3}\text{eV}}\big)^{1/2}\big(\frac{m_a}{10^{-4}\text{eV}}\big)}{\big(\frac{\epsilon}{10}\big)^2\big(\frac{m_a}{10^{-4}\text{eV}}\big)^2
+\big(\frac{\sqrt{3}\sigma}{10^{-3}\text{eV}}\big)^2}\Big)
\Big(\frac{\rho_a}{0.3\rm GeVcm^{-3}}\Big)^{1/2}
\Big(\frac{\delta t_{ob}}{1000\,\text{s}}\Big)^{1/2}	
\end{eqnarray}

Below, we have presented the figure where the boundaries between
SN ratio $I/I_n\times \sqrt{\delta \omega \delta t_{ob}/2\pi}>1$ and the ratio $<1$ are shown in $g_{a\gamma\gamma}-m_a$ plane. 
It shows that for larger axion masses, increasing the electrical conductivity is beneficial for the detection.

\begin{figure}[htp]
\centering
\includegraphics[width=0.60\hsize]{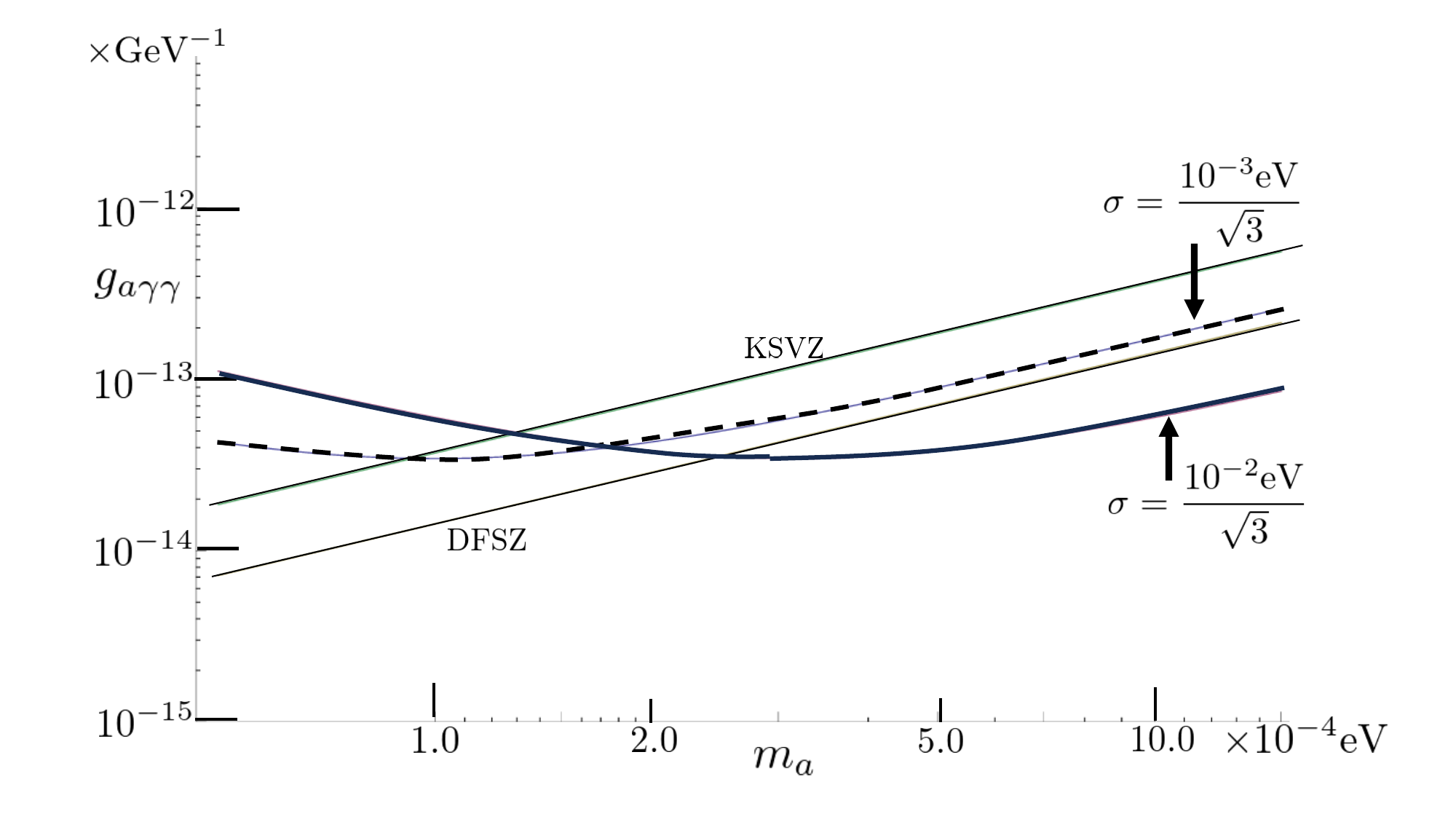}
\caption{Boundaries above which SN ratio $I/I_n\times \sqrt{\delta \omega \delta t_{ob}/2\pi}$ is larger than $1$ 
for two types of conductivity  $\sigma=10^{-3}\text{eV}/\sqrt{3}$ ( dashed ) 
and  $\sigma=10^{-2}\text{eV}/\sqrt{3}$ ( thick ), when $L=100\text{cm}$, $R=80\text{cm}$
and $B_0=7$T.
The ratio increases more as $g_{a\gamma\gamma}$ is bigger above the boundaries.}
\label{cylinder}
\end{figure}

\vspace{0.1cm}
We should mention a distinctive feature of our proposal for axion like particle ( ALP ) detection.
When the mass of the axion is extremely small $m_a\sim 10^{-14}$eV, 
it is detectable for even small sample with $L=10$cm and $R=1$cm by taking small
conductivity $\sigma\sim 10^{-13}\text{eV}$: The SN ratio is given by $\sim 10$.
That is, ALPs with extremely small mass are detectable 
using sample with extremely small conductivity being of the order of the mass, without the use of
large superconducting magnets to incorporate the sample. 
It can be understood by noting that the ratio $I/I_n\times \sqrt{\delta \omega \delta t_{ob}/2\pi}$ scales such as $\lambda^{-1/2}$
for $m_a\to m_a\lambda$ and $\sigma \to \sigma \lambda$. Thus, smaller $\lambda$ leads to larger ratio.

\section{conclusion}

In a strong magnetic field, axions generate an electric current in a cylindrical semiconductor sample. This current consists of two components: one localized near the surface due to conventional electromagnetic induction, and another that flows throughout the bulk of the semiconductor. The presence of this bulk current is a distinctive feature of axion-induced electromagnetism. In semiconductors, whose electrical conductivity is lower than that of metals but significantly higher than that of insulators, the bulk current becomes the dominant contribution.

In this paper,
we have proposed a method for axion detection with a cylindrical sample choosing
appropriate small conductivity $(\sigma = 10^{-3}\text{-} 10^{-2}\text{eV})$, which is
sensitive to microwave signals in the frequency range 
$m_a/2\pi = 24\text{-}240\text{GHz}$. With such conductivity, the axion induces large electric current $I$ in the bulk of the cylinder. It causes a high signal-to-noise ratio of the current $I\sqrt{\delta \omega \delta t_{ob}/2\pi}/I_n>1$ with the observation time $\delta t_{ob}=10^3$second. 
Therefore, the method demonstrates the feasibility of detecting dark matter axions ( or ALPs ) under realistic experimental conditions. It is also effective for the search of dark photon.

\vspace{0.1cm}




\end{document}